\documentclass[aps,prd,showpacs,notitlepage,nofootinbib,superscriptaddress,floatfix,showkeys,twocolumn,10pt]{revtex4-2}
\usepackage[colorlinks,citecolor=blue,urlcolor=blue,linkcolor=blue]{hyperref}
\usepackage[utf8]{inputenc}
\usepackage{blindtext}
\usepackage{amsmath,amssymb}
\usepackage{float}
\usepackage{microtype}
\usepackage{marginnote}
\usepackage{graphicx}
\usepackage{bm}
\usepackage{latexsym}
\usepackage{epsfig}
\usepackage{psfrag}
\usepackage{placeins}
\usepackage{color}
\usepackage[dvipsnames]{xcolor}
\usepackage{subfigure}
\usepackage{graphicx} 
\usepackage{amssymb}
\usepackage{amsmath}
\usepackage{bm}
\usepackage{latexsym}
\usepackage{epsfig}
\usepackage{psfrag}
\usepackage[normalem]{ulem}
\usepackage{textcomp}
\usepackage{color}
\usepackage{pstricks}
\usepackage[utf8]{inputenc}
\usepackage{url}
\usepackage{comment}
\usepackage{braket}
\usepackage{mathrsfs}
\usepackage{cleveref}
\usepackage{commath}

\usepackage{booktabs}
\usepackage{multirow}

\def\bea{\begin{eqnarray}}
\def\eea{\end{eqnarray}}
\def\nn{\nonumber}

\def\f{\frac}
\def\l{\left}
\def\r{\right}
\def\d{{\rm d}}
\def\Mpl{M_{_{\mathrm{Pl}}}}
\def\mpcinv{{\rm Mpc}^{-1}}

\def\cR{{\mathcal R}}
\def\cRk{\mathcal{R}_{\bm{k}}}
\def\vx{{\bm{x}}}
\def\vk{{\bm{k}}}
\def\vq{{\bm{q}}}
\def\ps{\mathcal{P}_{_{\mathrm{S}}}}

\def\stps{\mathscr{P}_{_{\rm h}}}

\def\barpow1d{\mathcal{P}^{1\rm D}_{\mathrm{b}}}

\def\ogw{\Omega_{_{\rm GW}}}

\newcommand{\IR}{\text{irr}}

\def\Dd{\delta^{(3)}}
\def\sph{\mathscr{P}_{_{\mathrm{h}}}}

\allowdisplaybreaks

\begin{document}

\title{Scalar-induced gravitational waves from coherent initial states}
\author{Dipayan Mukherjee}
\email{E-mail: dipayanmkh@gmail.com}
\affiliation{Raman Research Institute, C.~V.~Raman Avenue, Sadashivanagar, Bengaluru 560080, India}
\author{H.~V.~Ragavendra}
\email{E-mail: ragavendra.hv@pd.infn.it}
\affiliation{Dipartimento di Fisica e Astronomia “Galileo Galilei”, Università degli Studi di Padova, Via Marzolo 8, I-35131 Padova, Italy}
\affiliation{Istituto Nazionale di Fisica Nucleare, Sezione di Padova, Via Marzolo 8, I-35131 Padova, Italy}
\author{Shiv K.~Sethi}
\email{E-mail: sethi@rri.res.in} 
\affiliation{Raman Research Institute, C.~V.~Raman Avenue, Sadashivanagar, Bengaluru 560080, India}

\begin{abstract}
We investigate the impact of statistical inhomogeneity and anisotropy in primordial scalar perturbations on the scalar-induced gravitational waves (SIGW). Assuming inflationary quantum fluctuations originate from a coherent state, the resulting primordial scalar perturbations acquire a non-zero space-dependent mean, violating statistical homogeneity, statistical isotropy, and parity. As a consequence of statistical inhomogeneities, SIGW acquires distinct scale-dependent features in its correlation function. Statistical anisotropies further lead to possible parity violation and correlation between different polarization modes in the tensor perturbations. Therefore, detection of these signatures in the stochastic gravitational wave background would offer probes to the statistical nature of primordial scalar perturbations beyond the scales accessible to CMB observations.
\end{abstract}

\maketitle

\section{Introduction}

The cosmic microwave background (CMB) remains the most precise probe of the early universe, providing detailed constraints on inflationary physics and cosmological parameters~\cite{Planck2015XX,aghanim.akrami.ea20,Planck2018X}. However, CMB observations are limited to the largest scales --- typically beyond megaparsecs --- leaving small-scale physics largely unconstrained. 
In contrast, information encoded in cosmological gravitational waves (GW) remains unconstrained, particularly at much smaller scales, thus providing a window into the small-scale regime complementary to CMB~\cite{Dom_nech_2020, Hajkarim_2020, Caprini_2018, Domenech_2021, Inomata_2019a, Inomata_2019, Moore_2014, Ananda_2007, Baumann_2007, Cai:2020ovp, Cai:2023ykr}. With the advent of current and proposed GW experiments, the prospect of exploring the early universe at small scales seems increasingly tangible~\cite{Seto_2001, Lentati_2015, LISA_2017, Yagi_2011, LISA_2017, Reitze2019,Agazie_2023,Aggarwal_2021,EPTA_2023, PPTA_2023}.

One of the basic mechanisms for generating a stochastic background of GW is through tensor perturbations sourced by scalar fluctuations at the second order in perturbation theory~\cite{Tomita_1967, Matarrese_1993, Matarrese:1993zf,Matarrese:1996pp,Matarrese:1997ay,
Ananda_2007, Baumann_2007, Bartolo:2007vp, Guzzetti:2016mkm,
Kohri_2018, Espinosa_2018,Ragavendra:2020sop,Ragavendra_2021,Domenech_2021,
Ragavendra_2022,Adshead_2021,Ragavendra:2023ret,Perna_2024,Addazi:2024qcv,Wang:2018yql}.
These scalar-induced gravitational waves (SIGW) are appealing from a theoretical standpoint, as their production is inevitable and does not rely upon specifics of any model. Furthermore, since SIGW directly couple to the scalar modes, they also encode information about scalar perturbations, possibly at small scales that are otherwise inaccessible to electromagnetic probes. Notably, SIGW potentially observable today can probe modes that exited the horizon during very late stages of inflation, long after those accessible by CMB observation.

CMB observations infer a nearly scale-invariant primordial scalar power spectrum of amplitude $A_{\rm s} \simeq 10^{-9}$, which produces SIGW with amplitudes too weak to be detected by current or planned observations over large scales. 
However over smaller scales, the scalar power can deviate significantly from near-scale invariance and may lead to signals of SIGW detectable by probes over corresponding frequencies.
Such SIGW signals thus require physical processes that enhance power at small scales. Different mechanisms have been explored in the literature that could inject excess power at small scales (refer for a review~\cite{Domenech_2021}).

An intriguing source of enhancement in SIGW signal can be the deviation of the cosmological perturbations from statistical homogeneity and isotropy.
While the scalar perturbations are typically assumed to be statistically homogeneous and isotropic, deviations from such assumptions are not completely ruled out~\cite{Schwarz_2016, planck18_VII, Vielva_2010, Inoue_2006,Vielva_2004, Akrami_2014, Adhikari_2014, Gordon_2007, Hansen_2004, Land_2005,Schwarz_2004, de_Oliveira_Costa_2004, Kothari_2016, Rath_2015, Ackerman_2007, Souradeep_2006,Bartolo_2012, Gordon_2007}. Especially at scales that remain largely unexplored by CMB observations, scalar perturbations may deviate from statistical homogeneity and isotropy~\cite{Ragavendra:2024qpj}. Such primordial statistical inhomogeneities can contribute to the scalar power spectrum at small scales, subsequently enhancing the signal of SIGW to within the sensitivity of observational probes.

In this paper, we investigate the impact of statistical inhomogeneity and anisotropy in primordial scalar perturbations on SIGW. Specifically, we consider a scenario where quantum fluctuations in the inflaton field originate in a coherent state instead of the Bunch-Davies vacuum~\cite{Kundu_2014,Kundu2011,Mondal2024,Tejerina-Perez:2024opu,Ragavendra:2024qpj}. For a detailed discussion on the viability and consequence of such a choice of initial quantum states, see~\cite{Ragavendra:2024qpj} and references therein. Quantum fluctuations in a coherent state generally acquire a non-zero one-point function, resulting in a space-dependent mean for the primordial scalar perturbations. The presence of this \emph{primordial mean} violates the usual assumptions of statistical homogeneity and isotropy of the N-point correlation functions of scalar perturbations. However, the perturbations remain Gaussian. When these primordial fluctuations re-enter the horizon after the end of inflation, they induce tensor modes due to interactions at the second order of perturbation theory. As a result, the tensor perturbations sourced by these scalar modes inherit their statistical features, leading to distinctive signatures in their N-point correlation functions.

The presence of a statistically inhomogeneous primordial scalar mean introduces a one-point function, an induced mean, for the tensor perturbations.
Further it modifies the two-point correlation of the induced tensors. Beyond the homogeneous-isotropic power spectrum, this inhomogeneity generates reducible and partially reducible contributions. These contributions enhance the overall spectral density of SIGW. Crucially, the spatial profile of the primordial scalar mean determines the scale dependence of this enhancement. Localized features in the mean can produce a sharp peak in the SIGW's spectral density at specific scales while leaving larger and smaller scales unaffected. Detection of such scale-dependent enhancements in the stochastic GW background could constrain the structure of the primordial mean and therefore, the violation of statistical homogeneity of the primordial perturbations. 

Moreover, analysis of the anisotropic structure of the GW, especially the cross-correlation between unequal wavenumbers ($k\neq k'$), shall inform about the violation of statistical isotropy of the primordial perturbations. 
Besides, a statistically anisotropic primordial mean introduces preferred directions in the scalar perturbations. These directions couple to the polarization tensors of the induced tensor modes and break the symmetry between left- and right-handed polarization modes --- of the stochastic gravitational wave background. These scalar anisotropies can also induce cross-correlations between different polarization modes that are absent in the  homogeneous-isotropic scenarios. Consequently, detecting parity-violating signatures and primordial cross-correlation between polarizations in GW  also constrains statistical anisotropy in the primordial scalar perturbations.

It is also conceivable that the primary tensor fluctuations themselves originate from a coherent state, rather than the standard Bunch-Davies vacuum, resulting in inhomogeneities and anisotropies in the GW background. However, there is less physical motivation for excited initial states in the tensor sector compared to the scalar sector. While excitation in scalar perturbations might naturally occur from features in the inflaton potential or its interaction with other scalar fields~\cite{Armendariz_Picon_2007,Ragavendra_2021,Fumagalli:2021mpc}, primary tensor modes are sourced mainly by the metric perturbations. In the absence of additional tensor degrees of freedom or scalar-tensor couplings, there are no analogous interaction mechanism that could excite the tensor sector. 

This paper is organized as follows. We begin with a brief review of coherent states as the initial condition for inflationary perturbations in Sec.~\ref{sec:coherent_states}, along the lines of~\cite{Ragavendra:2024qpj}. In Sec.~\ref{sec:SIGWs}, we arrive at the scalar-induced tensor modes from the primordial scalar perturbations. We demonstrate how the primordial inhomogeneity in the scalar perturbations enhances the spectral density of SIGW in Sec.~\ref{sec:spectral_density}. In Sec.~\ref{sec:chirality}, we show that primordial scalar anisotropy leads to parity violation in SIGW and induces cross-correlations between different polarization modes. We conclude with a summary and discussion in Sec.~\ref{sec:conclusion}.

As to the notations used, $\eta$ denotes conformal time, $\vx$ is the comoving distance coordinate, $k$ is the comoving wavenumber.
We denote the reduced Planck mass $\Mpl = 1/\sqrt{8\pi G}$, setting $\hbar=c=1$.


\section{Coherent initial state}
\label{sec:coherent_states}
We assume the background dynamics of inflation to be that of a {\it slow-roll} model where a scalar field slowly rolls down a smooth potential. 
For the purpose of our calculation, such a background behavior is well described by a nearly constant Hubble parameter $H$ and small positive values of the first and second slow-roll parameters $\epsilon_1$ and $\epsilon_2$~\cite{Liddle:1994dx,Riotto:2002yw,Baumann2009}.
Over such a background we study the behavior of the scalar and tensor perturbations.

\subsection{Scalar perturbations}

We quantize the gauge-invariant, comoving curvature perturbation $\cR(\eta, \vx)$, 
using the annihilation and creation operators, $a_\vk$ and $a_\vk^\dagger$
as~\cite{Stewart:1993bc,Martin:2004um,Sriramkumar:2009kg}
\begin{eqnarray}
\cR(\eta, \vx) &=& \int \f{\d^3\vk}{(2\pi)^{3/2}} \cRk(\eta)e^{i \vk \cdot \vx}\,,\\
&=& \int \f{\d^3\vk}{(2\pi)^{3/2}}
\big[a_\vk \, f_k(\eta) \, e^{i \vk \cdot \vx} + 
a_{\vk}^\dagger \, f_k^\ast(\eta) \, e^{-i \vk \cdot \vx}\big]\,, \nn \\
\label{eq:scalar_pert}
\end{eqnarray}
where $f_k(\eta)$ is the mode function for a given wavenumber $k$. 
The creation and annihilation operators satisfy the standard commutation relation
\begin{equation}
[a_{\vk_1},a^{\dagger}_{-\vk_2}] = \delta^{(3)}(\vk_1+\vk_2)\,.
\end{equation}

Crucially, we evolve the perturbations from a coherent initial state instead of the Bunch-Davies vacuum.
The coherent state is defined as the eigenstate of the annihilation 
operator~\cite{Loudon:2000,Kundu2011,Kundu_2014,Mondal2024,Ragavendra:2024qpj}
\begin{eqnarray}
a_\vk \ket{C_\vk} &=& {\rm C}(\vk) \vert {C_\vk} \rangle\,,
\end{eqnarray}
where $C(\vk)$ is the eigenfunction for a given wavevector $\vk$. 
The coherent state is described by the function $C(\vk)=\vert C(\vk) \vert e^{i\theta_k}$, where $\theta_k$ is the related phase.
The number states $\ket{n_\vk}$ are created by $a^\dagger_\vk$ as 
$a^\dagger_\vk\ket{n_\vk}=\sqrt{n_\vk+1}\ket{n_\vk+1}$ 
and the lowest state 
$\ket{0}$ is the vacuum that is annihilated by $a_\vk$ as $a_\vk\ket{0}=0$.
The explicit expression of $\ket{C_\vk}$ in terms of the number states 
$\ket{n_\vk}$ is
\begin{eqnarray}
\ket{C_\vk} &=& e^{-\f{\vert C(\vk)\vert^2}{2}} \sum_{n=0}^\infty C^n(\vk) 
\f{\ket{n_\vk}}{\sqrt{n!}}\,.
\end{eqnarray}
The operator $D(\vk)$ that generates $\ket{C_\vk}$ from the vacuum $\ket{0}$ 
is called the coherent state displacement operator, whose form is~\cite{Loudon:2000,Mondal2024}
\begin{eqnarray}
D(\vk) &=& e^{-\f{\vert C(\vk) \vert^2}{2}} \sum_{n=0}^\infty 
C^n(\vk)\f{(a^{\dagger}_\vk)^n}{n!}\,,\\
&=& \exp\left({C(\vk) a^\dagger_\vk-\f{\vert C(\vk)\vert^2}{2}}\right)\,, \\
&=& \exp\left[C(\vk)a^\dagger_\vk - C^\ast(\vk) a_\vk\right]\,,
\end{eqnarray}
so that $D(\vk)\ket{0}=\ket{C_\vk}$. 

As to the mode function, $f_k(\eta)$, its evolution is governed by the 
slow-roll evolution of the background.
Using the behavior of such $f_k(\eta)$ in the super-Hubble regime, we may 
express the primordial mean acquired by $\cR_\vk$ in the coherent state 
as~\cite{Ragavendra:2024qpj}
\begin{eqnarray}
\langle \cR_\vk \rangle &=& i\l[\f{2\pi^2}{k^3}\ps(k)\r]^{1/2}\alpha(\vk)\,,
\label{eq:rave}
\end{eqnarray} 
where we define $\alpha(\vk) \equiv {\rm C}(\vk) - {\rm C}^\ast(-\vk)$ and
$\ps(k) \simeq H^2/(8\pi^2\Mpl^2\epsilon_1)$.
The function $\alpha(\vk)$ is complex and obeys the property $\alpha^\ast(-\vk)=-\alpha(\vk)$ to preserve the reality of $\braket{\cR(\vx)}$. 
The structure of the one-point function in real space is 
\begin{eqnarray}
\braket{\cR(\vx)} &=& \f{i}{2\sqrt{\pi}} \int \d^3 \vk
\sqrt{\f{\ps(k)}{k^3}} \, \alpha(\vk) \, e^{i\vk\cdot\vx}\,.
\label{eq:rx}
\end{eqnarray}
The homogeneity and isotropy are clearly violated at the perturbative level
due to the non-zero, space-dependent form of the {\it primordial mean} acquired
by the scalar perturbations.
The parity of the system is also violated as
\begin{eqnarray}
\braket{\cR(\vx)}-\braket{\cR(-\vx)} &=& \f{-1}{\sqrt{\pi}} \int \d^3 \vk
\sqrt{\f{\ps(k)}{k^3}}\, \nn \\
& &\times \Re[\alpha(\vk)] \, \sin(\vk\cdot\vx)\,.
\end{eqnarray}
Hence we obtain a distribution of $\braket{\cR(\vx)}$ that is inhomogeneous, anisotropic and has a parity-violating component.\footnote{This is an extension to the analysis in~\cite{Ragavendra:2024qpj}, where we had considered $\alpha(\vk)$ to be purely imaginary and thus $\braket{\cR(\vx)}$ to be parity invariant.}

The two-point function of $\cRk$ is then
\begin{eqnarray}
\langle \cR_{\vk_1} \cR_{\vk_2} \rangle &=& \f{2\pi^2}{(k_1k_2)^{3/2}}
\ps(k_1)\left[\delta^{(3)}(\bm k_1 + \bm k_2)\right. \nn \\ 
& & - \left.\alpha(\vk_1)\alpha(\vk_2) \sqrt{\f{\ps(k_2)}{\ps(k_1)}}\right]\,.
\label{eq:R-2ptfn}
\end{eqnarray}
We do not include the effects of any higher order interactions on $\cR$ and hence
there is no non-Gaussianity included.
Therefore, all the higher order correlations of $\cR$ shall be expressible in 
terms of its one-point and two-point correlations, i.e. using $\alpha(\vk)$ and $\ps(k)$.


\subsection{Tensor perturbations}

We turn to the tensor perturbations $\gamma_{ij}(\eta, \vx)$ and express them in Fourier space as~\cite{Caprini_2018, Maggiore_GW2_2018}
\begin{eqnarray}
\gamma_{ij}(\eta, \vx) &=& \sum_{\lambda} \int \f{\d^3\vk}{(2\pi)^{3/2}}
{\bm e}_{ij}^\lambda(\vk) \gamma^\lambda_\vk(\eta)e^{i\vk \cdot \vx}\,, \\
&=& \sum_{\lambda} \int \f{\d^3\vk}{(2\pi)^{3/2}}
{\bm e}_{ij}^\lambda(\vk)\,
\big[b^\lambda_\vk \, g_k^\lambda(\eta) \, e^{i \vk \cdot \vx} \nn \\
& & +\,b_{\vk}^{\lambda\,\dagger} \, g_k^{\lambda\,\ast}(\eta) \, 
e^{-i \vk \cdot \vx}\big]\,,
\label{eq:tensor_pert}
\end{eqnarray}
where ${\bm e}_{ij}^\lambda(\vk)$ is the polarization tensor for a given 
polarization index $\lambda$. The index $\lambda$ can be chosen to be $+,\,\times$
states or left and right circular polarization states.
The operators $b_\vk^\lambda$ and $b_\vk^{\lambda\,\dagger}$ are the 
annihilation and creation operators per polarization, obeying the commutation 
relation
\begin{equation}
[b_{\vk_1}^{\lambda_1},b^{\dagger\,\lambda_2}_{-\vk_2}] = \delta^{(3)}(\vk_1+\vk_2)\,\delta^{\lambda_1\lambda_2}\,.
\end{equation}

The tensor mode function $g^\lambda_k(\eta)$ is obtained from the corresponding
equation of motion solved over the slow-roll inflationary background.
Unlike the scalars, we evolve these tensor perturbations from the Bunch-Davies vacuum $\ket{0}$ during inflation. 
While the tensor perturbations can themselves be excited, we restrict our analysis to the case of tensors evolving from the vacuum state.
This is done to avoid any confusion between the effects on tensors imprinted by scalars evolving from coherent states and possible features due to tensors evolving from an excited initial state.
Therefore the two-point function of $\gamma^\lambda_\vk$ in the super-Hubble regime is~\cite{Caprini_2018, Maggiore_GW2_2018}
\begin{eqnarray}
\braket{\gamma^{\lambda_1}_{\vk_1}\gamma^{\lambda_2}_{\vk_2}} &=& 
\left(\f{2}{k^3}\right) \f{H^2}{\Mpl^2} \delta^{(3)}(\vk_1+\vk_2)
\delta^{\lambda_1\lambda_2}.
\end{eqnarray}

We shall then inspect the secondary tensors that are sourced by the scalar perturbations during radiation dominated epoch.

\section{Scalar-induced tensor perturbations}
\label{sec:SIGWs}

We shall compute the tensor perturbations that are induced by the scalars at the second-order, say $h^\lambda_{ij}(\eta,\vx)$, post inflation during the radiation-dominated era~\cite{Tomita_1967,Matarrese_1993,Baumann_2007,Ananda_2007,Kohri_2018,Domenech_2021}. 
Starting from a perturbed FLRW metric of the form
\begin{align}
\d s^2
= a(\eta)^2 
& \left[ -(1 + 2 \Phi) \d \eta^2 
\right.
\nonumber\\
& \left.
+ \left( (1 - 2 \Phi) \delta_{ij} + \frac{1}{2} h_{ij} \right) \d x^i \d x^j
\right],
\end{align}
the Fourier mode for such scalar-induced tensor perturbations can be written as~\cite{Kohri_2018,Espinosa_2018,Adshead_2021,Perna_2024,Ragavendra:2020sop,Ragavendra_2022}
\begin{align}
  h_\vk^\lambda(\eta)
  = 4 \int \frac{\d^3 \bm{q}}{(2 \pi)^{3/2}}
  Q^\lambda(\vk, \vq) I (|\vk - \vq|, q, \eta)
  \mathcal{R}_{\bm{q}} \mathcal{R}_{\bm{k-q} },
\end{align}
where the factor due to polarization tensor $Q^\lambda$ is given by
\begin{equation}
Q^\lambda(\vk,\vq) = {\bm e}^\lambda_{ij}(\vk) q^i q^j\,.
\end{equation}
It has the properties of
\begin{eqnarray}
Q^\lambda(-\vk,\vq) &=& {Q^\lambda}^\ast(\vk,\vq)\,, \\
Q^\lambda(\vk,-\vq) &=& Q^\lambda(\vk,\vq)\,,
\end{eqnarray}
in any given basis of polarization.

The coordinate representation of the polarization factor reveals further symmetries which will be useful for the discussion on chirality of the secondary tensor modes. 
If we align the vector $\bm{k}$ along the $\hat{\bm{z}}$ axis, $Q^\lambda$ in the $(+, \times)$ basis becomes
\begin{align}
\label{eq:Q^pc}
    Q^{+, \times}(\bm{k}, \bm{q})
    = \frac{1}{\sqrt{2}} q^2 \sin^2 \theta 
    \begin{cases}
        \cos 2 \phi & \text{ for }+\\
        \sin 2 \phi & \text{ for }\times
    \end{cases},
\end{align}
while in the helical basis
\begin{align}
\label{eq:Q^LR}
    Q^{L, R}(\bm{k}, \bm{q}) 
   = \frac{1}{2} q^2 \sin^2 \theta \exp(\pm 2 i \phi),
\end{align}
where $\theta$, $\phi$ are the polar and azimuthal angles of $\bm{q}$, respectively. Therefore $|Q^L|^2 = |Q^R|^2 = q^4 \sin^4 \theta/4$ has no azimuthal dependence, whereas the cross terms $Q^L Q^{R*} = (Q^{L*} Q^R)^* = q^4 \sin^4 \theta \exp(4 i \phi)/4$ depend on $\phi$.

The time evolution of $h^\lambda(\vk,\eta)$ is captured in the factor $I (|\vk - \vq|, q, \eta)$ which is given by~\cite{Domenech_2021, Adshead_2021, Kohri_2018,  Espinosa_2018}
\begin{align}
    I(p, q, \eta)
    =
    \int_{\eta_0}^{\eta} \d \tilde{\eta}
    G_{\bm{k}} (\eta, \tilde{\eta}) \frac{a(\tilde{\eta})}{a(\eta)}
    \mathfrak{f}(p, q, \tilde{\eta}),
\end{align}
where $G_{\bm{k}} (\eta, \tilde{\eta}) $ is the Green function that satisfies
\begin{align}
    G''_{\bm{k}} (\eta, \tilde{\eta}) + \left( k^2 - \frac{a''(\eta)}{a(\eta)}\right)  G_{\bm{k}} (\eta, \tilde{\eta}) 
    = \delta( \eta - \tilde{\eta}),
\end{align}
and $\mathfrak{f}(p, q, \tilde{\eta})$ is sourced by scalar perturbations
\begin{align}
  \mathfrak{f}(p, q, \eta) 
  =& \frac{3(1 + w)}{(5 + 3w)^2} 
  \left[ 2(5 + 3w)\Phi(p\eta)\Phi(q\eta)
    \right. \nonumber \\
  &\quad \left.
  + \eta^2(1 + 3w)^2\Phi'(p\eta)\Phi'(q\eta) 
      \right. \nonumber \\
  &\quad \left.
  + 2\eta(1 + 3w)\left( \Phi(p\eta)\Phi'(q\eta) + \Phi'(p\eta)\Phi(q\eta) \right) \right],
\end{align}
where $w$ is the background equation of state parameter. Note that $I(p, q, \eta)$ is symmetric under the exchange $p \leftrightarrow q$; this feature will allow us to simplify the expressions of correlation functions significantly.

Since the scalar perturbations in our scenario have a non-trivial primordial mean, 
they shall evidently have corresponding imprints on the structure of correlations 
of the induced tensor perturbations. So, we shall examine the structure of the
one-point and two-point correlations inherited by the scalar-induced tensor 
perturbations. 

\subsection{One-point function}

The one-point function of the scalar-induced secondary tensor mode shall be
\begin{align}
\braket{h^\lambda_\vk(\eta)}
= 4 \int \frac{\d^3 \bm{q}}{(2 \pi)^{3/2}}
Q^\lambda(\bm{k}, \bm{q}) I (|\bm{k} - \bm{q}|, q, \eta)
\Braket{\mathcal{R}_{\bm{q}} \mathcal{R}_{\bm{k-q} }}.
\end{align}
With the primordial mean of $\braket{\cR(\vx)}$, the two-point function in the integrand becomes
\begin{align}
\braket{h^\lambda_\vk(\eta)} &= 4 \int \frac{\d^3 \bm{q}}{(2 \pi)^{3/2}}
Q^\lambda(\bm{k}, \bm{q}) I (|\bm{k} - \bm{q}|, q, \eta) \nn \\
&\quad \times \left[\f{2\pi^2}{q^3}\ps(q) \Dd(\bm{k})
+ \Braket{\mathcal{R}_{\bm{q}}} \Braket{\mathcal{R}_{\bm{k-q} }} \right]
\end{align}
The first term in the square braces contributes to the background (only for $\vk = \bm 0$), and so we are left with
\begin{eqnarray}
\braket{h^\lambda_\vk(\eta)}  &=& -8\pi^2 \int \frac{\d^3 \bm{q}}{(2 \pi)^{3/2}}
Q^\lambda(\bm{k}, \bm{q}) I (|\bm{k} - \bm{q}|, q, \eta) \nn \\
& & \times \sqrt{\f{\ps(q)\ps(\vert \vk-\vq\vert)}{q^3\vert\vk-\vq\vert^3}}\,\alpha(\vq)\alpha(\vk-\vq)\,.
\label{eq:h_one_p_fun}
\end{eqnarray}
Thus tensor perturbation acquires an one-point function due to the source at the second-order having a non-vanishing mean.
The reality condition of tensor perturbations is still preserved as we can see 
that $\braket{h^\lambda_{-\vk}}=\braket{h^{\lambda}_\vk}^\ast$.
Further, if we write the explicit forms of $\braket{h^\lambda_\vk}$ in helical
basis, we shall find that
\begin{eqnarray}
\braket{h^R_\vk(\eta)} &=& -8\pi^2 \int \f{\d^3\vq}{(2\pi)^{3/2}} 
I(|\vk - \vq|, q, \eta) Q^R(\vk,\vq) \nn \\ & & \times 
\sqrt{\f{\ps(q)\ps(\vert \vk-\vq\vert)}{q^3\vert\vk-\vq\vert^3}}\,
\alpha(\vq)\alpha(\vk-\vq)\,.
\label{eq:h_R} \\
\braket{h^L_\vk(\eta)} &=& -8\pi^2 \int \f{\d^3\vq}{(2\pi)^{3/2}} 
I(|\vk - \vq|, q, \eta) Q^L(\vk,\vq) \nn \\ & & \times
\sqrt{\f{\ps(q)\ps(\vert \vk-\vq\vert)}{q^3\vert\vk-\vq\vert^3}}\,
\alpha(\vq)\alpha(\vk-\vq)\,.
\label{eq:h_L}
\end{eqnarray}
We can see that $\braket{h^{R,L}_{-\vk}} = \braket{h^{R,L}_\vk}$ only if
$[\alpha(\vq)\alpha(\vk-\vq)]^\ast = \alpha(\vq)\alpha(\vk-\vq)$, i.e. if
$\alpha(\vk)$ is a fully real or a fully imaginary function. 
In case of $\alpha(\vk)$ being a general complex function, we shall have
$\braket{h^\lambda_{-\vk}} \neq \braket{h^\lambda_\vk}$ indicating violation of 
parity in tensors, inherited from the sourcing scalars.
Thus we have a non-trivial one-point function for the scalar-induced tensor 
perturbations, that are inhomogeneous, anisotropic and violating parity.


\subsection{Two-point function}

The two-point correlation of $h^\lambda_\vk$ at a given $\eta$, in terms of
the four-point function of the scalar perturbations is
\begin{align}
  \label{eq:1}
  \braket{h^{\lambda_1}_{\vk_1}(\eta) h^{\lambda_2}_{\vk_2}(\eta)}
  =& 16 \iint \frac{\d^3 \bm{q}_1}{(2 \pi)^{3/2}}
     \frac{\d^3 \bm{q}_2}{(2 \pi)^{3/2}}
     Q^{\lambda_1}(\bm{k}_1, \bm{q}_1)
     Q^{\lambda_2}(\bm{k}_2, \bm{q}_2) \nonumber\\
   &\quad \times
     I (|\bm{k}_1 - \bm{q}_1|, q_1, \eta)
     I (|\bm{k}_2 - \bm{q}_2|, q_2, \eta)\nonumber\\
   &\quad \times
     \Braket{
     \mathcal{R}_{\bm{q}_1} \mathcal{R}_{\bm{k}_1-\bm{q}_1 }
     \mathcal{R}_{\bm{q}_2} \mathcal{R}_{\bm{k}_2-\bm{q}_2 }
     }.
\end{align}
Since we do not assume any non-Gaussianity for the primordial scalar perturbation
$\cR$, the four-point function in the integrand above decomposes as\footnote{Here, we use $\cR_1$ as a shorthand for $\cR_{\vk_1}$ and so on.}
\begin{subequations}
  \begin{align}
    &\Braket{\mathcal{R}_1 \mathcal{R}_2 \mathcal{R}_3 \mathcal{R}_4} = 
       \Braket{\mathcal{R}_1 \mathcal{R}_2 \mathcal{R}_3 \mathcal{R}_4}_\IR\\
    &+
       \Braket{\mathcal{R}_1} 
       \Braket{\mathcal{R}_2 \mathcal{R}_3 \mathcal{R}_4}_\IR
       + \text{3 permutations}\\
    &+
       \Braket{\mathcal{R}_1 \mathcal{R}_2}_\IR
       \Braket{\mathcal{R}_3 \mathcal{R}_4}_\IR
       + \text{2 permutations}
       \label{eq:r4-r2}\\
    &+
       \Braket{\mathcal{R}_1} \Braket{\mathcal{R}_2}
       \Braket{\mathcal{R}_3 \mathcal{R}_4}_\IR
       + \text{5 permutations}
       \label{eq:r4-r1-r2}\\
    &+ 
       \Braket{\mathcal{R}_1} 
       \Braket{\mathcal{R}_2} 
       \Braket{\mathcal{R}_3} 
       \Braket{\mathcal{R}_4} 
       \label{eq:r4-r1}
  \end{align}
\end{subequations}
The terms in the first and second line of RHS are zero, since there are no irreducible three- and four-point correlations for Gaussian perturbations. 
The terms in the third line~\eqref{eq:r4-r2} are from the four-point function reduced in terms of irreducible two-point functions. 
Of the three permutations between these terms only two survive. 
One term contains $\delta^{(3)}(\vk_1)\delta^{(3)}(\vk_2)$ and so does not contribute to non-zero values of $k_1$ or $k_2$. 
These two terms are the statistically homogeneous and isotropic contributions independent of the primordial mean. 
Next, in the fourth line~\eqref{eq:r4-r1-r2}, there are terms from four-point function reducible to products of one-point and irreducible two-point functions.
Once again, of the six terms, two shall contain $\delta^{(3)}(\vk_1)$ and $\delta^{(3)}(\vk_2)$ which do not contribute at non-zero wavenumbers.
The last line~\eqref{eq:r4-r1} is the term that is completely reducible in terms of one-point function.

Collecting the surviving terms, the two-point function of the induced tensor perturbations can be expressed as
\footnote{The functions $f$ and $g$ should not be confused with the mode functions $f_k$ and $g_k$ defined in Eqs.~\eqref{eq:scalar_pert} and~\eqref{eq:tensor_pert}.}
\begin{align}
\braket{h^{\lambda_1}_{\vk_1} h^{\lambda_2}_{\vk_2}} = &
f^{\lambda_1}(k_1 , \eta)
\delta^{\lambda_1\lambda_2}\Dd(\vk_1 + \vk_2) 
+ g^{\lambda_1\lambda_2}(\vk_1 , \vk_2 , \eta)
\nn \\
& +\braket{h^{\lambda_1}_{\vk_1}}\braket{h^{\lambda_2}_{\vk_2}}\,.
\label{eq:h1h2}
\end{align}
The first term in the RHS above encapsulates contribution from~\eqref{eq:r4-r2}. 
The presence of $\Dd(\vk_1+\vk_2)$ indicates the homogeneous and isotropic nature of this term.
The Kronecker delta $\delta^{\lambda_1\lambda_2}$ implies parity.
The explicit form of $f_{\lambda}(k{, \eta})$ is given by
\begin{align}
  f^{\lambda}(k {, \eta}) = 32 & (2\pi^2)^2\int \f{\d^3 \vq}{(2\pi)^3}
  \vert Q^{\lambda}(\bm{k}, \bm{q})\vert^2 I^2(|\vk - \vq|, q, \eta) \nn\\
  & \times \f{\ps(q)\ps(|\vk - \vq|)}{q^3\vert\vk-\vq\vert^3}\,.
\label{eq:f-term}
\end{align}
The second term of Eq.~\eqref{eq:h1h2} captures the contribution from~\eqref{eq:r4-r1-r2}. 
Note that it is neither completely reducible nor irreducible, as it contains
terms that mix one-point and two-point functions.
The form of $g^{\lambda_1\lambda_2}(\bm{k}_1, \bm{k}_2 {, \eta})$ is given by
\begin{align}
  g^{\lambda_1\lambda_2}(\bm{k}_1, \bm{k}_2 {, \eta})
  =& -64 (2\pi^2)^2\int \frac{\d^3 \bm{q}}{(2 \pi)^3}
     Q^{\lambda_1}(\bm{k}_1, \bm{q})
     Q^{\lambda_2}(\bm{k}_2, \bm{q}) \nonumber\\
   &\times  
     I (|\bm{k}_1 - \bm{q}|, q, \eta) 
     I (|\bm{k}_2 + \bm{q}|, q, \eta) \nonumber\\
   &\times 
     \sqrt{\f{\ps(\vert\vk_1-\vq\vert)\ps(\vert\vk_2+\vq\vert)}{\vert\vk_1-\vq\vert^3\vert\vk_2+\vq\vert^3}}
     \f{\ps(q)}{q^3} \nn \\
    &\times 
    \alpha(\vk_1-\vq)\alpha(\vk_2+\vq)\,.
\label{eq:g-term}
\end{align}
Note that we have used the symmetries of the functions $I$ and $Q$ to derive these quantities (see Sec.~\ref{sec:SIGWs}).
The last term of Eq.~\eqref{eq:h1h2}, fully reducible in terms of 
$\braket{h^\lambda_\vk}$, evidently comes from the four one-point functions in~\eqref{eq:r4-r1}.
Its explicit expression shall be
\begin{eqnarray}
\braket{h^{\lambda_1}_{\vk_1}}
\braket{h^{\lambda_2}_{\vk_2}} &=& 16\,(2\pi^2)^2 
\Bigg[\int \frac{\d^3 \vq_1}{(2 \pi)^{3/2}}
    I(|\vk_1 -\vq_1|, q_1, \eta) \nn \\
    & & \times Q^{\lambda_1}(\vk_1,\vq_1) 
        \sqrt{\f{\ps(q_1)\ps(\vert \vk_1-\vq_1\vert)}{q_1^3\vert\vk_1-\vq_1\vert^3}}\nn \\
    & & \times \alpha(\vq_1)\alpha(\vk_1-\vq_1)\Bigg] \nn \\
    & & \times \Bigg[\int \frac{\d^3 \vq_2}{(2 \pi)^{3/2}}
    Q^{\lambda_2}(\vk_2, \vq_2) I (|\vk_2 - \vq_2|, q_2, \eta) \nn \\
    & & \times \sqrt{\f{\ps(q_2)\ps(\vert \vk_2-\vq_2\vert)}{q^3\vert\vk_2-\vq_2\vert^3}}\,\alpha(\vq_2)\alpha(\vk_2-\vq_2)\Bigg]\,. \nn \\
\label{eq:h-term}
\end{eqnarray}


\section{Spectral density of secondary GW}
\label{sec:spectral_density}

Having known the one-point and two-point functions of the induced tensor perturbations, we shall proceed to compute the spectral density of the energy density of the associated secondary GW, in terms of the dimensionless parameter
$\ogw(k,\eta)$.
For this computation, we shall consider the case where $\alpha(\vk) = \alpha(k)$, i.e.~the primordial mean is inhomogeneous but isotropic and preserves parity.
This case can be viewed as the contribution to the monopole of the two-point correlation arising from the spatially-averaged part of $\alpha(\vk)$ (cf.~similar discussion regarding spectral distortion in~\cite{Ragavendra:2024qpj}).

Notice that when $\alpha(\vk)=\alpha(k)$, the condition to preserve reality of $\braket{\cR(\vx)}$, $\alpha(-\vk) = -\alpha^\ast(\vk)$ simply becomes
$\alpha(k) = -\alpha^\ast(k)$, enforcing $\alpha$ to be fully imaginary.
Hence, preserving isotropy restores parity in $\braket{\cR(\vx)}$ and 
$h^\lambda_\vk$.
Further, the one-point function of $\braket{h^\lambda_\vk(\eta)}$ simply vanishes
in this case, due to azimuthal integration.
This can be understood by substituting the factor due to 
polarization tensor $Q^\lambda(\vk,\vq)$ given in Eq.~\eqref{eq:Q^pc} 
or~\eqref{eq:Q^LR} in $\braket{h^\lambda_\vk(\eta)}$ expressed in 
Eq.~\eqref{eq:h_one_p_fun}.
With $\alpha(q)\alpha(\vert \vk-\vq\vert)$ being function of just $q$ and 
the polar angle $\theta$,
the only dependence on the azimuthal angle $\phi$ arises in terms of 
$\exp(\pm2i\phi)$ from $Q^\lambda(\vk,\vq)$ in the integrand, 
which vanishes upon integration over $\phi$.
Therefore, the effect of $\alpha(k)$ on secondary GW shall be solely through the
partially reducible term $g^{\lambda_1\lambda_2}(\vk_1,\vk_2{, \eta})$ in the two-point function [cf.~Eq.~\eqref{eq:h1h2}].

In order compute the inhomogeneous and homogeneous contributions in the two-point function, we adopt the technique used in~\cite{Ragavendra:2024qpj}: we integrate the dimensionless combination of the two-point correlation 
$(k_1 k_2)^{3/2} \Braket{h^{\lambda_1}(\vk_1) h^{\lambda_2}(\vk_2)}/(2 \pi^2)$
over a small volume in Fourier space to arrive at an equivalent of the power spectrum, denoted as $\stps^{\lambda_1\lambda_2}(\vk_1,\vk_2,\eta)$.

To obtain $\stps^{\lambda_1\lambda_2}(\vk_1,\vk_2,\eta)$, with $\vk_2 = -\vk_1$, 
we integrate over $\vk_2$ over a volume of $k_1^3\Delta^3$ in the direction of 
$-\hat \vk_1$, with $\Delta$ being a small dimensionless parameter.
This gives us $\stps^{\lambda_1\lambda_2}(\vk_1,-\vk_1,\eta)$ to be
\begin{align}
\stps^{\lambda_1\lambda_2}(\vk,-\vk,\eta) 
& = \delta^{\lambda_1\lambda_2}
\f{k^3}{2\pi^2}\,\left[f^{\lambda_1}(k{, \eta}) + \right.
\nn\\
&\left.\qquad k^3\Delta^3\,g^{\lambda_1\lambda_1}(\vk,-\vk {, \eta})\right]\,
\label{eq:phk-k}
\end{align}
The assumption in performing the integration is that $g(\vk,-\vk)$ does not vary much in $k$-space over the small volume of $k^3\Delta^3$, so that it can be simply extracted out 
of the integration.
Hence, the inhomogeneous part of the spectral density seemingly scales as $\Delta^3$.
But we should note that this scaling is not valid for arbitrary values of $\Delta$. 
If $\Delta$ is larger than the characteristic scale of variation in $g(\vk_1,\vk_2)$, 
the expression of Eq.~\eqref{eq:phk-k} is not true anymore and one has to perform the integration over 
$\vk_2$ accounting for the variation of $g(\vk_1,\vk_2)$ within the integration volume.
For our analysis, we shall work with value of $\Delta$ which satisfies this requirement.
For an alternative yet equivalent derivation of spectral density, that does not rely on the smallness parameter $\Delta$, refer Appendix~\ref{app:spectral_density}.

Substituting the explicit forms of the functions in $\stps^{\lambda_1\lambda_2}(\vk,-\vk,\eta)$ given above, we get
\begin{align}
\stps^{\lambda_1\lambda_2}(\vk,-\vk,\eta) 
&= 64\pi^2\,\delta^{\lambda_1\lambda_2}\,k^3
\int \frac{\d^3 \bm{q}}{(2 \pi)^3} \vert Q^\lambda(\bm{k}, \bm{q})\vert^2 
\nn \\ & \times
I^2(|\bm{k} - \bm{q}|, q, \eta) 
\f{\ps(q)}{q^3} \f{\ps(|\vk - \vq|)}{|\bm{k} - \bm{q}|^3} \nn \\
& \times \left[ 1 + 2 \Delta^3 k^3\,\vert\alpha(\vert\bm{k} - \bm{q}\vert)\vert^2\right]\,.
\label{eq:phk-k-alpha}
\end{align}
Notice that setting $\vk_2 = -\vk_1$ brings about $\delta^{\lambda_1\lambda_2}$
in the second term too.
We can identify the first term in $\stps^{\lambda \lambda}(\vk,-\vk,\eta)$ as the power spectrum of the induced tensor-modes in the homogeneous-isotropic scenario~\cite{Domenech_2021, Caprini_2018}, while the second term is the correction due to statistical inhomogeneity.

For $\vk_2 \neq -\vk_1$, we integrate the quantity over $\vk_1'$ around $\vk_1$
and over $\vk_2'$ around $\vk_2$, with the same smallness parameter $\Delta^3$
determining the volume of integration and obtain the average between the two
integrations as
\begin{eqnarray}
\stps^{\lambda_1\lambda_2}(\vk_1,\vk_2 {, \eta}) &=& \left(\f{k_1^3+k_2^3}{2}\right)
\Delta^3\,\f{(k_1k_2)^{3/2}}{2\pi^2}
\nn\\
& & \times\,g^{\lambda_1\lambda_2}(\vk_1,\vk_2{, \eta})\, \nn \\
\\
&=& -64\pi^2 \left(k_1^3+k_2^3\right) \Delta^3\,(k_1k_2)^{3/2} \nn \\
& & \times \int \frac{\d^3 \bm{q}}{(2 \pi)^3}
Q^{\lambda_1}(\bm{k}_1, \bm{q})
Q^{\lambda_2}(\bm{k}_2, \bm{q}) \nonumber\\
& &\times I (|\bm{k}_1 - \bm{q}|, q, \eta) 
I (|\bm{k}_2 + \bm{q}|, q, \eta) \nonumber\\
& &\times \f{\ps(q)}{q^3} 
\sqrt{\f{\ps(\vert\vk_1-\vq\vert)\ps(\vert\vk_2+\vq\vert)}{\vert\vk_1-\vq\vert^3\vert\vk_2+\vq\vert^3}} \nn \\
& &\times \alpha(\vert\vk_1-\vq\vert)\alpha(\vert\vk_2+\vq\vert)\,.
\label{eq:phk1k2}
\end{eqnarray}
Note that in this term, there is no guarantee of $\delta^{\lambda_1\lambda_2}$. 
Hence, there can be cross-correlation between different polarizations in 
$\stps^{\lambda_1\lambda_2}(\vk_1,\vk_2 {, \eta})$ with $\vk_2 \neq -\vk_1$.
We shall examine the nature of correlations between different polarizations in 
different types of terms in a later section.

Having computed the two-point correlation of the induced tensor perturbations,
we shall proceed to compute the spectral density of the associated
secondary GW today, in terms of the dimensionless parameter $\ogw^0$.
Recall that the total energy density associated with gravitational waves can be written in terms of the two-point function of the tensor modes as~\cite{Caprini_2018,Domenech_2021}
\begin{align}
    \rho_{\mathrm{GW}}
     = \frac{\Mpl^2}{16 a^2 (\eta)} \Braket{{h}'_{ij}(\eta, \bm{x} ){h}'_{ij}(\eta, \bm{x} ) }.
    \end{align}
The corresponding spectral density is defined as 
\begin{align}
  \ogw
  = \frac{1}{\rho_c} 
     \frac{\d \rho_{\mathrm{GW}}}{\d \ln k},
\end{align}
where $\rho_c$ is the critical density,  $\rho_c \equiv 3 \Mpl^2 H^2$.
In the statistically homogeneous-isotropic scenario, the spectral density relates to the power spectrum of the tensor modes $\mathcal{P}^\lambda_{\mathrm h}$ as~\cite{Domenech_2021, Adshead_2021, Kohri_2018, Ragavendra_2022, Perna_2024, Espinosa_2018}
\begin{align}
  \ogw(k {, \eta}) =  \frac{k^2}{48 a^2 H^2}
     \sum_\lambda \mathcal{P}_{\mathrm h}^\lambda(k, \eta),
\end{align}
using the sub-Hubble behavior of tensor modes propagating freely, $h'(k) \simeq k h(k)$. 
In the presence of statistical inhomogeneity, we may extend the above relation to
\begin{align}
  \ogw(\vk_1, \vk_2 {, \eta})
  = \f{k^2}{48 a^2 H^2} \sum_\lambda\stps^{\lambda \lambda} (\bm{k}_1, \bm{k}_2, \eta).
\end{align}
Assuming that the secondary gravitational waves are induced in the radiation dominated era, we may then evaluate the quantity today as~\cite{Kohri_2018,Espinosa_2018}
\begin{align}
h^2 \ogw^0(\bm k, -\bm k) \simeq 1.38 \times 10^{-5} \,
\ogw(\bm k, -\bm k)\,,
\label{eq:ogw_today}
\end{align}
where $h$ is defined in terms of Hubble parameter today $H_0$ 
as $H_0=100\,h\,{\rm km}\,{\rm s}^{-1}\mpcinv$. 
{Note that the spectral density becomes time-independent in this scenario, as we shall see later.}

\subsection{Parameterization of the primordial mean}
For further analysis, we will use the following general parameterization of the scalar primordial mean
\begin{align}
  \alpha(\bm{k}) = -\frac{i}{k^{3/2}} \alpha_0 F(k) \beta(\hat {\bm{k}}),
\label{eq:alpha}
\end{align} 
where the dimensionless functions $F(k)$ and $\beta(\hat {\bm{k}})$ capture the magnitude and angular dependence of the primordial mean, respectively, while the dimensionless parameter $\alpha_0$ sets the overall amplitude. Since we preserve isotropy and focus just on inhomogeneity of $\alpha(\vk)$ in this section, we set $\beta(\hat {\bm{k}}) = 1$.

We are specifically interested in generation of enhancement in $\ogw$.
Before proceeding with the numerical evaluation of $\ogw$, we should note that 
the perturbativity of $\braket{\cR}$ imposes a constraint on $\alpha_0$.
Recall that we obtain $\ps(k) \simeq H^2/(8\pi^2\Mpl^2\epsilon_1)$ under slow-roll 
approximation.
However, $\ps(k)$ can deviate significantly from this estimate due to features in the 
potential and/or non-trivial background dynamics.
If we assume $\ps(k)$ to be scale-invariant $\ps(k)=A_{\rm s}$ for simplicity 
and substitute Eq.~\eqref{eq:alpha} in Eq.~\eqref{eq:rx}, then 
$\braket{\cR} < 1 \Rightarrow \alpha_0\sqrt{A_{\rm s}} < 1$. 
Hence, for a given value of $A_{\rm s}$, we may not be able to set arbitrary values 
of $\alpha_0$, but ensure $\alpha_0 < 1/\sqrt{A_{\rm s}}$\,.
While the amplitude $A_{\rm s}$ is well-constrained on large scales, it can be
enhanced over small scales that remain largely unconstrained (see for instance~\cite{Ragavendra:2023ret,Ragavendra:2024yfp}).
This shall tighten the bound on $\alpha_0$.
On the other hand, if $A_{\rm s}$ remains small over small scales, then $\alpha_0$
can reach reasonably large values.
We shall explore the interplay of these parameters in the amplitude of $\ogw$.

For simplicity of analysis, let us assume that both $\ps(k)$ and $\alpha(k)$ are 
enhanced around the same scale $k_0$.
To focus on the scale dependence of $\ogw$ due to $\alpha(k)$, let us also assume 
that $\ps(k)$ is enhanced over a sufficiently broad range around $k_0$, such 
that it can be approximated as $\ps(k)=A_{\rm s}$ in the computation.
We model the scale-dependence of $\alpha(k)$ as
\begin{align}
F(k) = \f{1}{\sqrt{2\pi\Delta_f^2}}
\exp \left[- \frac{1}{2 \Delta_f^2} \ln^2 \left( \frac{k}{k_0} \right)\right],
\label{eq:log-normal}
\end{align}
with $\Delta_f$ determining the sharpness of the peak.
The strength of GW at $k_0$, if the inhomogeneous contribution dominates, 
shall scale as $\ogw \propto \alpha_0^2 A_{\rm s}^2$ [cf.~Eq.~\eqref{eq:phk-k-alpha}].
If we then impose the constraint on $\alpha_0$ by fixing the combination 
$\alpha_0\sqrt{A_{\rm s}}$ as a constant value smaller than unity, 
then $\ogw \propto A_{\rm s}$. 
This behavior is in contrast to the homogeneous contribution which scales as 
$\ogw \propto A_{\rm s}^2$.


With this understanding of the parametric dependences, we now proceed with the exact numerical evaluation of the diagonal part of $\ogw$ ($\vk_2 = -\vk_1$) as
\begin{align}
  \ogw(\vk,-\vk {, \eta})
  = \f{k^2}{48 a^2 H^2} \sum_\lambda\sph^{\lambda \lambda} (\bm{k}, -\bm{k}, \eta).
\end{align}
To evaluate $\sph^{\lambda \lambda}(\bm{k}, -\bm{k} {, \eta})$ numerically, it is convenient to align the wave vector $\bm{k}$ along the $\hat{\bm{q}}_{\bm z}$-direction. We also define the following dimensionless variables involved in the integration 
\begin{align}
\label{eq:mu-v}
  \mu \equiv   \hat{\bm k} \cdot \hat{\bm q} = \cos \theta ,\quad
  v \equiv \frac{q}{k},
\end{align}
Recasting the integrals in terms of these variables and summing over the two polarization modes ($+,\times$ or $R,L$), we get 
\begin{align}
  \label{eq:P_h}
  \sum_\lambda
  \sph^{\lambda \lambda} (\bm{k}, -\bm{k} {, \eta}) & = 8
    \int_0^{\infty} \d v \int_{-1}^{1}\d \mu\  \left( 1 - \mu^2 \right)^2
    \frac{v^3}{u^3}    \nn \\
  &\times
    \tilde{I}^2(u,v,x)
    \ps(uk)  \ps(vk)\nn \\
    &\times 
    \Bigg[1 + 2 \frac{\Delta^3 \alpha_0^2}{2\pi\Delta_f^2 u^3} 
    \exp \left[-\frac{1}{ \Delta_f^2} \ln^2 \left( \frac{uk}{k_0} \right) \right]
    \Bigg]\,,\qquad
\end{align}
where we have introduced $ u(\mu, v) \equiv \sqrt{v^2 + 1 - 2 v \mu}$, $x \equiv k \eta$, and the kernel function $\tilde{I}(u, v, x) \equiv k^2 I(uk, vk, x/k)$~\cite{Espinosa_2018,Kohri_2018,Adshead_2021}.

In the following discussion, we assume that the secondary tensor modes are induced in the radiation-dominated era. In principle, the tensor modes can also be sourced during inflation, however this contributions are typically subdominant compared to those generated in the radiation era~\cite{Fumagalli:2021mpc,Baumann_2007}. 
The oscillation-averaged product of the kernel function in the radiation-dominated era is given by~\cite{Espinosa_2018,Kohri_2018,Adshead_2021}
\begin{align}
  &\tilde{I}(u_1, v_1, x \to \infty)
  \tilde{I}(u_2, v_2, x \to \infty)
  \nonumber \\
  =&
     \frac{1}{2x^2}
     I_A(u_1, v_1)
     I_A(u_2, v_2)
     \nonumber\\
   &\times
     \left(
     I_B(u_1, v_1)
     I_B(u_2, v_2)
     + \pi^2
     I_C(u_1, v_1)
     I_C(u_2, v_2)
     \right),
     \label{eq:I_tilde}
\end{align}
where
\begin{subequations}
  \begin{align}
    \tilde{I}_A (u,v)
    &=
      \frac{3 (u^2 + v^2 -3)}{4u^3 v^3}\\
    \tilde{I}_B (u,v)
    &= -4uv + (u^2 + v^2 -3) \ln \abs{ \frac{3 - (u+v)^2}{3 - (u-v)^2}}\\
    \tilde{I}_C (u,v)
    &= \left( u^2 + v^2 -3 \right) \Theta(v + u - \sqrt{3}).
  \end{align}
\end{subequations}
Owing to radiation domination, we set $ a H = 1/\eta$ in the above expression.
{Consequently, the spectral density  $ \ogw(\vk,-\vk, \eta)$ becomes time-independent, and we drop the explicit $\eta$ dependence in what follows.}
Under these assumptions, we numerically evaluate $\sum_\lambda \sph^{\lambda \lambda}(\vk, -\vk, \eta)$ and  hence $h^2\ogw^0(\vk,-\vk)$.

\begin{figure}
  \centering
    \subfigure[$h^2 \ogw^0(\vk, -\vk)$ vs.~wave numbers for different $A_{\rm s}$.]{
    \includegraphics[width=\columnwidth]{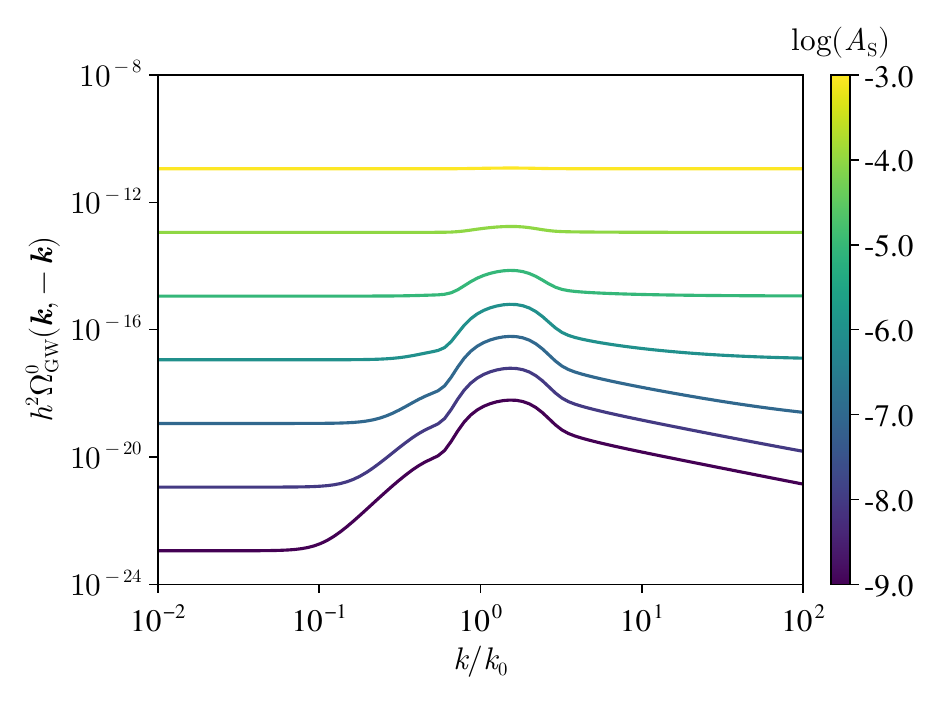}
    \label{fig:omega_by_k_vary_As}
  }
  \hfill
  \subfigure[$h^2 \ogw^0(\vk, -\vk)$ vs.~$A_{\rm s}$ at different wave numbers.]{
    \includegraphics[width=\columnwidth]{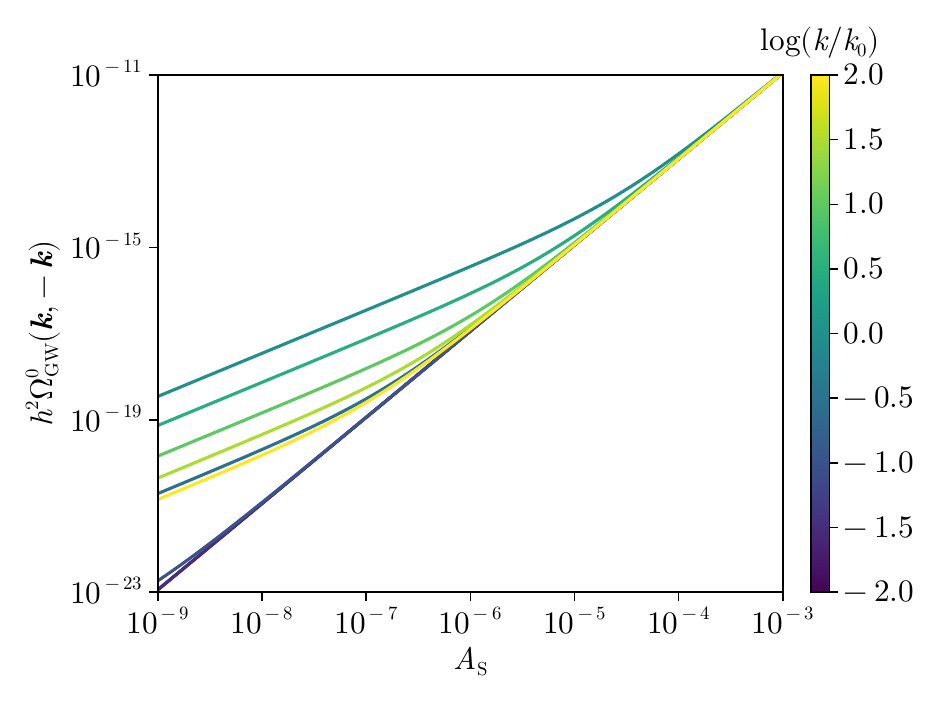}
    \label{fig:omega_by_As_vary_k}
  }
\caption{Enhancement in the spectral density of secondary gravitational waves induced by statistical inhomogeneity in the primordial scalar perturbations. Fig.~\ref{fig:omega_by_k_vary_As} shows $\ogw^0$ as a function of wave numbers for different $A_{\rm s}$.
While $\ogw^0$ increases with increasing $A_{\rm s}$, the relative enhancement due to the primordial mean decreases.
Fig.~\ref{fig:omega_by_As_vary_k} presents $\ogw^0$ as a function of $A_{\rm s}$ evaluated at different wave numbers near its peak.
For smaller $A_{\rm s}$ the peak of $\ogw^0$ varies linearly with $A_{\rm s}$, while it varies quadratically for larger $A_{\rm s}$.
We have set the model parameters to be $\ps(k) = A_{\rm s}$,
$\alpha_0=0.1/\sqrt{A_{\rm s}}$, $\Delta = 0.1$ and $\Delta_f = 0.2$.}
\label{fig:spectral-density}
\end{figure}

Figure~\ref{fig:spectral-density} shows the enhancement in $\ogw^0(\bm k, -\bm k)$ across a range of scales $k$ and $A_{\rm s}$.
We set $\alpha_0=0.1/\sqrt{A_{\rm s}}$ to ensure perturbativity of $\braket{\cR}$ discussed earlier.
Figure~\ref{fig:omega_by_k_vary_As} shows that the enhancement in $\ogw^0$ due to the primordial mean is most pronounced for smaller values of $A_{\rm s}$.
When $A_{\rm s}$ is comparable to its CMB-constrained value ($A_{\rm s} \sim 10^{-9}$), the spectral density peaks sharply as $k$ approaches $k_0$, with a maxima at $k \simeq 1.58\,k_0$.
Post the peak, for $k>k_0$, the spectral density falls with a relatively flatter slope.

Far away from $k_0$, the spectral density approaches a constant value consistent with that of the statistically homogeneous scenario, $ \ogw \simeq 0.82\,A_{\rm s}^2$~\cite{Kohri_2018}. Therefore, the effect of primordial mean in $\ogw$ is restricted to the vicinity of the scale where the mean peaks and $\ogw$ remains unaffected at scales which are much larger or smaller.
For larger values of $A_{\rm s}$, $\alpha_0$ decreases due to the constraint of $\alpha_0=0.1/\sqrt{A_{\rm s}}$.
As a result, although $\ogw^0$ increases with increasing $A_{\rm s}$, the relative enhancement in $\ogw^0$ due to the primordial mean is weakened.

Figure~\ref{fig:omega_by_As_vary_k} presents $\ogw^0$ as a function of $A_{\rm s}$ evaluated at different $k$ values near the peak. 
As argued earlier, for smaller values of $A_{\rm s}$, we find at $k_0$, $\ogw^0 \propto 
A_{\rm s}$. This suggests the dominance of the inhomogeneous contribution in this regime. 
As $A_{\rm s}$ increases, the fractional contribution from the inhomogeneous term becomes  subdominant, and so $\ogw^0 \propto A_{\rm s}^2$ consistent with the homogeneous scenario.
At scales away from $k_0$, $\ogw^0$ scales steeper than $\ogw^0 \propto A_{\rm s}$, indicating weaker contribution from $\alpha(k)$.
This exercise informs us that the enhancement in $\ogw$ can not be of significant values, such as $\ogw^0 \geq 10^{-15}$, if it is solely from $\alpha(k)$, and  likely to be a contribution of both $\alpha(k)$ and $\ps(k)$. 
Further at large values of $\ogw$, it is more likely to be dominated by the 
homogeneous contribution and the effect of $\alpha(k)$ may be discernible as features on the profile of $\ogw$.

We also compute the off-diagonal terms of $\ogw$ ($k_1\neq k_2$).
The spectral density in a general configuration of wave vectors $\ogw(\vk_1,\vk_2)$, can be expressed as
\begin{align}
  \ogw(\vk_1, \vk_2)
  = \frac{k^2}{48 a^2 H^2} \sum_\lambda\sph^{\lambda \lambda}(\bm{k}_1, \bm{k}_2, \eta).
\end{align}
For numerical evaluation, let us assume a simpler scenario where the wave vectors are antiparallel,
$\hat{\bm k}_1 = - \hat{\bm k}_2 = \hat{\bm q}_{\bm z}$. 
We consider the log-normal form of the primordial mean as before, (Eq.~\eqref{eq:log-normal}), and recall that we have set the anisotropic factor $\beta(\hat {\bm k}) = 1$. Using the convenient redefinitions
\begin{align}
  \mu 
  \equiv \hat{\bm k}_1 \cdot \hat{\bm q}  
  = - \hat{\bm k} _2 \cdot \hat{\bm q}
  = \cos \theta_ 
  ,\quad
  v \equiv \frac{q}{k},
\end{align}
 and summing over polarizations, we get
\begin{align}
  \sum_{\lambda} \stps^{\lambda \lambda}(\bm{k}_1, \bm{k}_2,\eta)
  &= 16 \frac{\Delta^3 \alpha_0^2}{2\pi\Delta_f^2} \left[  \frac{1}{2}\f{\left( k_1^{3} + k_2^3 \right)}{k_a^6} \left( k_1 k_2 \right)^{\frac{3}{2}} \right]
     \nonumber\\
  &\times
    \int_0^{\infty} \d v \int_{-1}^{1}\d \mu\ \frac{v^3}{u^3 w^3}
    \left( 1 - \mu^2 \right)^2
    \nonumber\\
  &\times
    \tilde{I}(u,v,x) \tilde{I}(w,v,x)
\nonumber\\
  &\times
    \ps(v k_a) \sqrt{\ps(u k_a) \ps (w k_a)}  
\nonumber\\
  &\times
    \exp \left[ - \frac{1}{2 \Delta_f^2} \ln^2 \left( \frac{u k_a}{k_0} \right)\right] 
\nn \\
    &\times
    \exp\left[ - \frac{1}{2 \Delta_f^2} \ln^2 \left( \frac{w k_a}{k_0} \right)\right]\,,
\end{align}
where,
\begin{align}
  k_a(k_1, k_2) &\equiv \frac{k_1 + k_2}{2},\\
  u(\mu, v) &\equiv \sqrt{v^2 + \frac{k_1^2}{k_a^2} - 2 \frac{k_1}{k_a} v \mu},\\
  w(\mu, v) &\equiv \sqrt{v^2 + \frac{k_2^2}{k_a^2} - 2 \frac{k_2}{k_a} v \mu}.
\end{align}
Utilizing these expressions, we numerically evaluate $\sum_\lambda\stps^{\lambda \lambda} (\bm{k}_1, \bm{k}_2, \eta)$ and obtain the associated $\ogw(\vk_1,\vk_2)$.

\begin{figure}
\centering
\includegraphics[width=\columnwidth]{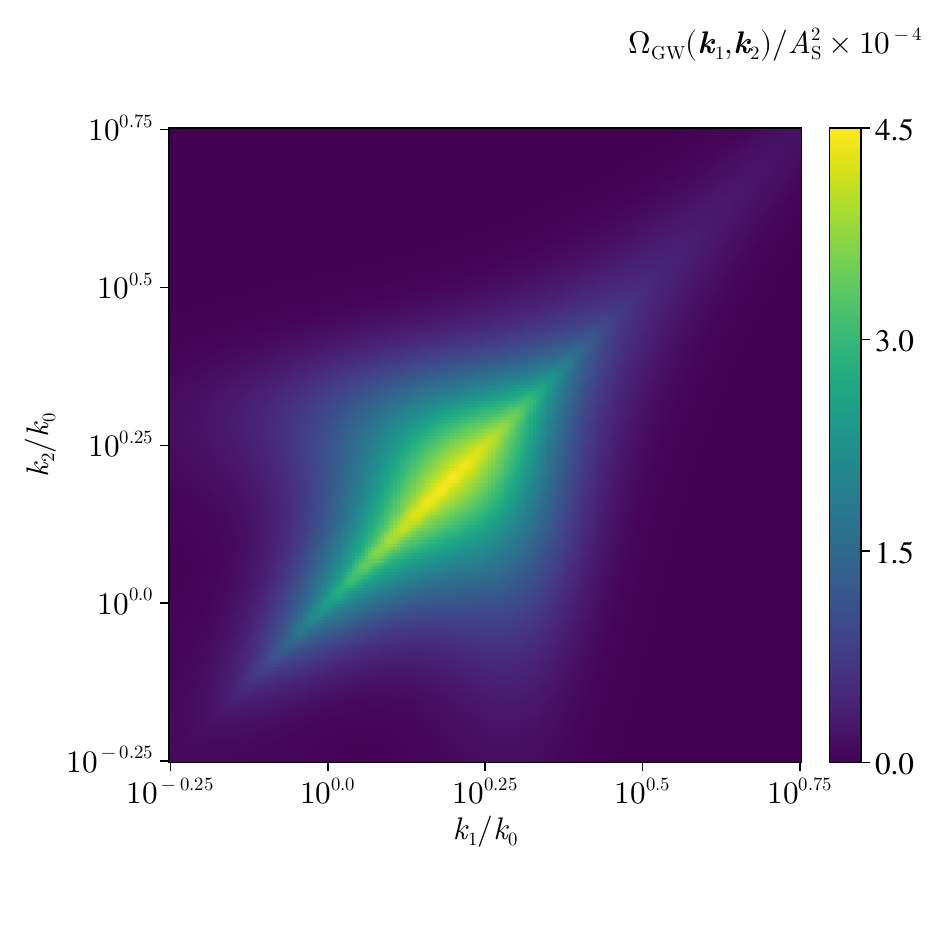}
\vskip -0.35in
\caption{Enhancement in the spectral density of secondary gravitational waves induced by statistical inhomogeneity in the primordial scalar perturbations.
The heatmap shows $\ogw( \bm k_1, \bm k_2)/ A_{\rm s}^2$ with respect to the wavenumbers $k_1, k_2$, where the wave vectors are antiparallel.
We have set $\ps(k) = A_{\rm s}$ with $A_{\rm s} = 10^{-9}$, other model parameters are same as in Fig.~\ref{fig:spectral-density}.
Although the amplitude in the diagonal part may be degenerate with the homogeneous 
contribution, or even dominated by it for higher values of $A_{\rm s}$ as in 
Fig.~\ref{fig:spectral-density}, we find the off-diagonal part directly determined by
the amplitude and shape of $\alpha(k)$.}
\label{fig:off-dia}
\end{figure}
Figure~\ref{fig:off-dia} shows the complete structure of $\ogw(\vk_1,\vk_2)$.
As we can see, $\ogw(\vk_1,\vk_2)$ peaks on the diagonal $k_1 = k_2$ and gradually 
vanishes at regions away from the diagonal line. 
The diagonal contribution arises from both the homogeneous and inhomogeneous 
components of $\stps^{\lambda_1\lambda_2}(\vk_1,\vk_2,\eta)$, whereas the 
off-diagonal contribution is solely from the inhomogeneous part arising from 
$\alpha(k)$.
The spectral density peaks around the scale of inhomogeneity $k_1 \approx k_2 \approx k_0$, and decays away from $k_0$.
It is interesting to note that the structure of $\ogw(\vk_1,\vk_2)$ is not 
symmetric about the line $\ln(k_2/(1.58\,k_0))=-\ln(k_1/(1.58\,k_0))$ in Fig.~\ref{fig:off-dia}. This asymmetric structure is induced by the 
kernel of integration arising from factors of $\tilde I$ and polarization factors
apart from $\alpha(k)$.
 
We should emphasize that we have worked with simple parametric choices for 
$\ps(k)$ and $\alpha(k)$ for illustrating these major effects in the behavior of $\ogw$.
More sophisticated functional forms of $\ps(k)$ and $\alpha(k)$, motivated 
by realistic inflationary models leading to features in them, may uncover 
further interesting effects in the structure of $\ogw$.


\subsection{Parameterization of the primordial mean: Dirac $\delta$ function}

As another example, we obtain the spectral density of the SIGW using a Dirac $\delta$ function for the scalar primordial mean. This exercise will reveal the analytical form of the enhancement in the SIGW signal due to the statistical inhomogeneities.  

Let us consider the inhomogeneous part of the polarization-summed $\stps^{\lambda \lambda}(\bm{k}, -\bm{k} {, \eta} )$ [cf.~second term of Eq.\eqref{eq:phk-k-alpha}]
\begin{align}
  \sum_\lambda {\stps^{\lambda \lambda}}_{\rm (ih)}(\bm{k}, -\bm{k} {, \eta})
 =
  &
    \frac{2^4}{\pi}
    k^3
    \int \d^3 \bm{q}
    \sum_\lambda Q_\lambda^2(\bm{k}, \bm{q}) 
    \nn\\
    \times&
    I^2(|\bm{k} - \bm{q}|, q, \eta)
    \frac{\ps(q)}{q^3}
    \frac{\ps(|\bm{k} - \bm{q}|)}{|\bm{k} - \bm{q}|^3}
     \nn\\
    \times&
     \Delta^3 k^3 
    \abs{\alpha(\bm{q})}^2
    ,
\end{align}
where we have performed the transformation $\bm{q} \to \bm{k} - \bm{q}$,
and used the symmetries of $Q_\lambda (\bm{k}, \bm{q})$ and
$I(p, q, \eta)$ for further simplifications.
Let us consider again the lognormal parametrization of $\alpha(\vk)$ [cf.~Eq.~\eqref{eq:log-normal}] and inspect the functional form of $F^2(k)$ appearing in the integrand of $\stps^{\lambda \lambda}(\vk, -\vk, \eta)$.
We take the limit of $\Delta_f \to 0$ which reduces $F^2(k)$ to be
\begin{align}
F^2(k) &= \delta\left( \ln\frac{k}{k_{0}} \right)
\label{eq:Dirac_delta_profile}
\end{align}
We proceed as before and use the notations introduced in the previous subsection [cf.~Eq.~\eqref{eq:mu-v}] to write
\begin{align}
  \sum_\lambda {\stps^{\lambda \lambda}}_{\rm (ih)}(\bm{k}, -\bm{k} {, \eta})
  &= 2^4 \alpha_0^2 \Delta^3
    \frac{k_0}{k}
    \int_{-1}^{1}\d \mu(1-{\mu}^2)^{2}
    \frac{1}{u_0^3}
    \nn\\
    &\times
    \mathcal{P}_{s}(u_0k)\mathcal{P}_{s}(k_0)
    \tilde{I}^2\left(u_0,v_0,x \right)
    \label{eq:delta_Ph}
\end{align}
where $v_0 \equiv k_0/k,\ u_0 \equiv u(v_0, \mu) = \sqrt{1 + v_0^2 - 2 \mu v_0}.$
We can perform the integration analytically in the asymptotic large- and small-scale limits, i.e., when $v_0 = k_0/k \gg 1$ and $k_0/k \ll 1$, which will reveal the analytical profiles of the rise and fall of the enhancement in the spectral density.
We consider $\ps(k) = A_{\rm s}$ as before.

\begin{figure}
\centering
\includegraphics[width=\columnwidth]{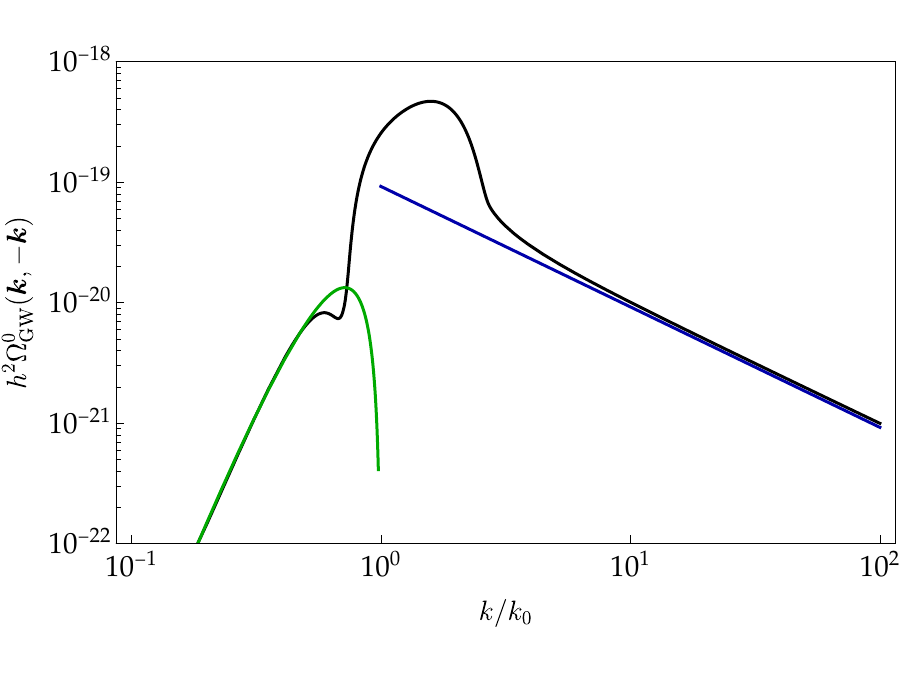}
\vskip -0.35in
\caption{
Enhancement in the spectral density of secondary gravitational waves induced by a scalar primordial mean with a Dirac-$\delta$ profile (cf.~Eq.~\eqref{eq:Dirac_delta_profile}).
The black curve shows inhomogeneous contribution to $\ogw^0(\bm{k}, -\bm{k})$ obtained from the exact numerical solution Eq.~\eqref{eq:delta_Ph}, while the green and blue curves show the  analytical solutions for small and large $k$ limits (Eq.~\eqref{eq:delta_Ph_small_k} and Eq.~\eqref{eq:delta_Ph_large_k}), respectively.
The enhancement in the spectral density rises $\propto (k/k_0)^6 \ln^2(k/k_0)$, followed by a more gradual fall $\propto (k/k_0)^{-1}$ after peaking near $k/k_0 = 1$.
We have set the model parameters to be $\ps(k) = A_{\rm s} = 10^{-9}$,
$\alpha_0=0.1/\sqrt{A_{\rm s}}$, and $\Delta = 0.1$.}
\label{fig:asymptotic}
\end{figure}

First consider the regime where $v_0 \gg 1 \Rightarrow k \ll k_0$, which implies $u_0 \approx v_0$. The function $\tilde{I}(u_0, v_0, x)$ can be approximated in this region as 
\begin{equation}
\tilde{I}^2(v_0, v_0, x) \approx \f{18}{k^2 \eta^2 v_0^4}\ln^2 v_0\,,
\end{equation}
[cf.~Eq.~\eqref{eq:I_tilde}]. 
This leads to the asymptotic expression 
\begin{align}
    \sum_\lambda {\stps^{\lambda \lambda}}_{\rm (ih)}(\bm{k}, -\bm{k} {, \eta}) 
    \approx &
    \frac{1536}{5 k^2 \eta^2} \alpha_0^2 \Delta^3 A_{\rm s}^2
    \left( \frac{k}{k_0}\right)^6 \ln^2 \left( \frac{k}{k_0} \right), 
    \nn \\
    \label{eq:delta_Ph_small_k}
\end{align}
for $k \ll k_0$.
In the regime where $v_0 \ll 1 \Rightarrow k \gg k_0$ (and so $u_0 \simeq 1$), the function $\tilde{I}(u_0, v_0, k)$ can be approximated as $\tilde{I}^2(u_0, v_0, k) \approx 2/{k^2 \eta^2}$, leading to the asymptotic form of
\begin{align}
    \sum_\lambda {\stps^{\lambda \lambda}}_{\rm (ih)}(\bm{k}, -\bm{k} {, \eta}) 
    \approx & 
    \frac{2^5}{k^2 \eta^2} \alpha_0^2 \Delta^3 A_{\rm s}^2 \frac{k_0}{k}\,,
    \label{eq:delta_Ph_large_k}
\end{align}
for  $k\gg k_0$.
Using these expressions, we obtain the asymptotic behaviors of the diagonal part of the spectral density $\ogw^0(\bm{k}, -\bm{k})$ [cf.~Eq.~\eqref{eq:ogw_today}]. Figure~\ref{fig:asymptotic} shows $\ogw^0(\bm{k}, -\bm{k})$ obtained from the exact numerical solution Eq.~\eqref{eq:delta_Ph} (black plot), the analytical solution for small $k$ Eq.~\eqref{eq:delta_Ph_small_k} (green plot), and the analytical solution for large $k$ Eq.~\eqref{eq:delta_Ph_large_k} (blue plot). As the analytical results indicate, the enhancement in the spectral density has a sharp rise $\ogw^0\propto (k/k_0)^6 \ln^2(k/k_0)$, followed by a more gradual decline $\ogw^0\propto (k/k_0)^{-1}$ after peaking around $k/k_0 = 1$. 


\section{Chirality in secondary GW}
\label{sec:chirality}

Due to the parity violating component of $\alpha(\vk)$, we can expect parity-odd
signature of chirality in the secondary GW induced.
As seen earlier, the anisotropic component of $\alpha(\vk)$ is responsible for
parity violation.
Further, $\alpha(\vk)$ also induces correlation between different polarization 
modes of the tensor perturbations [cf.~Eq.~\eqref{eq:phk1k2}], which are 
uncorrelated in the statistically homogeneous-isotropy scenario. 

We shall begin by inspecting the chirality of the one-point function of the
induced tensor perturbations $\braket{h^\lambda_\vk}$.
Let us retain a generic structure of the $\alpha(\vk)$ as given in
Eq.~\eqref{eq:alpha}.
We shall not assume any specific form of $F(k)$ and allow for a general angular 
dependence in the anisotropy inducing factor $\beta(\hat{\bm{k}})$, expanding it in 
terms of spherical harmonics as
\begin{align}
  \beta(\bm{k}) &= \sum_{\ell = 0}^{\infty} \sum_{m = -\ell}^{\ell} \beta_{\ell m} Y_{\ell m}(\hat{\bm{k}}).
\label{eq:beta}
\end{align}
Using the properties of $Y_{\ell m}$ and expressing it in terms of 
the associated Legendre polynomials $P_{\ell m}$ in Eq.~\eqref{eq:h_one_p_fun}, 
the left- and right-handed tensor one-point functions can be written as 
\begin{align}
\Braket{h_{\bm{k}}^{L,R}(\eta)} &= \frac{2\pi^2}{(2\pi)^{3/2}} \alpha_0^2 
    \sum_{\ell}^{\infty} \sum_{\ell' = 0}^{\infty} \sum_{m = -\ell}^{\ell} 
    \beta_{\ell m} \beta_{\ell' m'}
    \nn \\
  &\times 
    (-1)^{m}
    \sqrt{(2\ell+1)(2\ell'+1)}
    \sqrt{\frac{(\ell - m)! (\ell' - m')!}{(\ell + m)! (\ell' + m')!} } 
    \nonumber \\
  &\times 
    \int \d q \,q \int \d(\cos\theta) 
    \frac{\sqrt{\ps(p)\ps(q)}}{p^3} 
    F(p) F(q)
    \nonumber \\
  &\times 
    \sin^2 \theta 
    P_{\ell m}(\cos\theta) 
    P_{\ell' m'}\left( \frac{k - q\cos\theta}{p} \right)
    I(p, q, \eta),
\label{eq:hlr}
\end{align}
where the index $m'$ is related to $m$ as
\begin{align}
m' = \begin{cases}
-(m + 2) & \text{for } L\,, \\
-(m - 2) & \text{for } R\,,
\end{cases}
\end{align}
and $\bm{p} \equiv \bm{k} - \bm{q}$.
It is evident from the above selection rule that for a constant
$\beta(\hat{\bm{k}})$, i.e., where there are only monopole
contributions ($\ell, \ell'=0$) in $\beta_{\ell m}$, the tensor
one-point function identically vanishes for both left- and
right-handed modes. 
Therefore, the primordial mean of the scalar perturbation must violate 
statistical isotropy to induce a non-zero mean $\braket{h^\lambda_\vk}$ for 
the secondary tensor perturbations.

For a general $\beta(\hat{\vk})$, $\braket{h^\lambda_\vk}$ of 
different polarization modes $\lambda$ are different essentially due to
the selection rules differing as $m' = m\pm 2$. So, we have a parity-odd
one point function as
$\braket{h_{\bm{k}}^{L}(\eta)} \neq \braket{h_{\bm{k}}^{R}(\eta)}.$

As to the two-point function, for the terms with $\vk_2=-\vk_1$ we have
\begin{eqnarray}
\stps^{\lambda_1\lambda_2}(\vk,-\vk) &=& \f{k^3}{2\pi^2}\,
\big[f^{\lambda_1}(k)\delta^{\lambda_1\lambda_2} +
k^3\Delta^3g^{\lambda_1\lambda_2}(\vk,-\vk)\big] \nn \\ 
& & +\Delta^3\f{k^6}{2\pi^2} \braket{h^{\lambda_1}_\vk}
\braket{h^{\lambda_2}_\vk}^\ast,
\end{eqnarray}
which differs from Eq.~\eqref{eq:phk-k} by the second and third terms.
We may recall that for $\vk_2=-\vk_1$, we obtain $\stps^{\lambda_1\lambda_2} \propto \delta^{\lambda_1\lambda_2}$ when $\alpha(\vk)=\alpha(k)$. 
But the angular dependence included in $\alpha(\vk)$ enables 
$g^{\lambda_1\lambda_2}$ to acquire non-zero values for $\lambda_1\neq\lambda_2$. 
Further, the fully reducible part contributes due to non-zero $\braket{h^{\lambda}_\vk}$.
Interestingly, $g^{\lambda_1\lambda_2}$ does not contribute to the parity violation in the  $\vk_2=-\vk_1$ configuration, i.e. $g^{LL}=g^{RR}$.
This can be seen from Eq.~\eqref{eq:g-term},
\begin{align}
g^{LL, RR}(\bm{k}, -\bm{k}) = & 64\,(2 \pi^2)^2 \int \frac{\d^3 \bm{q}}{(2 \pi)^3}
I^2(|\bm{k} - \bm{q}|, q, \eta)\nn\\
& \times
\vert Q^{L,R}(\bm{k}, \bm{q}) \vert^2 
\frac{\ps(q)}{q^3}
\frac{\ps(|\bm{k} - \bm{q}|)}{|\bm{k} - \bm{q}|^3} \nn \\
& \times \abs{\alpha(\bm{k} - \bm{q})}^2\,,
\end{align}
and noting $|Q^{L}|^2 = |Q^R|^2 $ from Eq.~\eqref{eq:Q^LR}.
However, since $\braket{h^\lambda_\vk}$ is parity violating, we find that
\begin{align}
\stps^{LL}(\vk,-\vk)-\stps^{RR}(\vk,-\vk) =& \Delta^3\f{k^6}{2\pi^2}
\big[\vert\braket{h^L_\vk}\vert^2 - \vert\braket{h^R_\vk}\vert^2\big]\,,
\end{align}
and it is non-zero due to Eq.~\eqref{eq:hlr}.
We may quantify chirality in terms of spectral density, focussing on terms with 
$\vk_2=-\vk_1$ as
\begin{eqnarray}
\Pi_k &=& \f{\stps^{LL}(\vk,-\vk)-\stps^{RR}(\vk,-\vk)}{\stps^{LL}(\vk,-\vk)+\stps^{RR}(\vk,-\vk)}\,, \\
&=& \f{k^3\Delta^3 [\vert\braket{h^L_\vk} \vert^2 - \vert \braket{h^R_\vk} \vert^2]}
{\sum_{\lambda=L,R}\big[f^{\lambda}(k) + k^3\Delta^3\big(g^{\lambda\lambda}(\vk,-\vk) + \vert \braket{h^\lambda_\vk} \vert^2\big)\big]}\,.
\nn \\
\label{eq:chi}
\end{eqnarray}
Hence, we see that the chirality in the spectrum of secondary GW shall be directly
proportional to the difference between the squares of the one-point functions of the induced tensors in the two polarization states.
In case of the contribution to $\stps^{\lambda_1\lambda_2}$ due to the 
inhomogeneous and anisotropic terms $g^{\lambda\lambda}$ and 
$\vert\braket{h^\lambda_\vk}\vert^2$ being subdominant to the 
contribution from the homogeneous and isotropic part $f^\lambda(k)$, 
$\Pi_k$ can be approximated as
\begin{eqnarray}
\Pi_k &\simeq& \f{k^3\Delta^3[\vert\braket{h^L_\vk} \vert^2 - \vert\braket{h^R_\vk} \vert^2]}
{f^L(k) + f^R(k)}\,.
\end{eqnarray}

Moreover, we also obtain correlation between different polarization modes as 
\begin{eqnarray}
\stps^{LR}(\vk,-\vk) &=& \Delta^3\f{k^6}{2\pi^2} 
\big[g^{LR}(\vk,-\vk) + \braket{h^L_\vk}\braket{h^R_\vk}^\ast\big]\,,~~~~~ \\
&=& [\stps^{RL}(\vk,-\vk)]^\ast\,,
\end{eqnarray}
which arise from terms quadratic and quartic in $\alpha(\vk)$.

For the terms with $\bm{k}_1 \neq -\bm{k}_2$, we have  
$\sph^{\lambda_1, \lambda_2}(\bm{k}_1, \bm{k}_2)$ expressible as
\begin{eqnarray}
\stps^{\lambda_1\lambda_2}(\vk_1,\vk_2) &=& \left(\f{k_1^3+k_2^3}{2}\right)
\Delta^3\,\f{(k_1k_2)^{3/2}}{2\pi^2} \big[g^{\lambda_1\lambda_2}(\vk_1,\vk_2) \nn \\
& & + \braket{h^{\lambda_1}_{\vk_1}}\braket{h^{\lambda_2}_{\vk_2}}\big]\,,
\end{eqnarray}
Evidently, the left- and right-circular components in this configuration
are different $\sph^{LL}(\bm{k}_1, \bm{k}_2) \neq \sph^{RR}(\bm{k}_1, \bm{k}_2)$
and once again, we have non-vanishing correlation between different polarization 
modes $\sph^{LR}(\bm{k}_1, \bm{k}_2) \neq 0$.
We should point out that the non-vanishing values of correlations between 
different polarizations (such as $L$ and $R$ or equivalently $+$ and $\times$) 
is a signature of parity-violation occurring along with the violation of 
homogeneity and isotropy.
Hence, in case of degeneracy between our scenario of interest and other
mechanisms of parity-violation leading to similar $\Pi_k$,
such a correlation between polarizations can resolve the degeneracy.

\section{Conclusion}
\label{sec:conclusion}

We have examined the generation of scalar-induced GW arising in the scenario of 
primordial scalar perturbations evolving from a coherent initial state.
In this case,  the scalar perturbations possess a non-trivial primordial mean 
determined by $\alpha(\vk)$. 
In our previous work, we had shown that this leads to   the violation of  statistical homogeneity and 
isotropy of scalar perturbations at the perturbative level~\cite{Ragavendra:2024qpj}.
In this work,  we have shown  that, in addition,  the anisotropic structure of $\alpha(\vk)$ causes the violation of parity.

We  have computed the correlation functions that the tensor 
perturbations inherit from the scalar perturbations through interactions at the second
order of perturbation theory.
The tensor perturbations acquire a non-zero, scale-dependent one-point function
$\braket{h_\vk^\lambda}$, thus acquiring signatures of inhomogeneity, anisotropy 
and parity-violation.
We have computed the dimensionless spectral density of the tensor perturbations 
$\sph^{\lambda_1\lambda_2}(\vk_1,\vk_2)$ and the associated observable 
$\ogw^0(\vk_1,\vk_2)$.
Using a lognormal parameterization of $\alpha(k)$, we have illustrated that the
amplitude of $\ogw^0 \propto \vert\alpha(k)\vert^2$ can acquire  new features, especially for low amplitudes of $A_{\rm s}$ (Figure~\ref{fig:spectral-density}). 

Due to the nontrivial nature of the one-point function $\alpha(\vk)$, $\sph^{\lambda_1\lambda_2}(\vk_1,\vk_2)$ displays other interesting properties.
In particular, We find that there are non-vanishing correlations of $\sph^{\lambda_1\lambda_2}(\vk_1,\vk_2)$
between $\vk_2 \neq \vk_1$, which are unique signatures of statistical inhomogeneity and anisotropy (Figure~\ref{fig:off-dia}). In the previous paper, we have shown that the two- and three-point correlation functions of scalar perturbations also reveal similar signatures ~\cite{Ragavendra:2024qpj}.

We have then investigated the implications of parity violation. 
It is shown  that the anisotropic structure of $\alpha(\vk)$ leads to chiral imbalance 
$\sph^{LL}(\vk_1,\vk_2) \neq \sph^{RR}(\vk_1,\vk_2)$ for terms with $\vk_2=-\vk_1$ as well as
$\vk_2\neq-\vk_1$.
A dimensionless quantifier of chirality $\Pi_k$, for the terms with $\vk_2=-\vk_1$,
turns out to be proportional to the difference between $\vert \braket{h_\vk^\lambda}\vert^2$
in different polarizations [cf.~Eq.~\eqref{eq:chi}].
Moreover, we observe non-vanishing cross-correlation between modes of different 
polarizations, i.e. non-zero $\sph^{\lambda_1\lambda_2}(\vk_1,\vk_2)$ with
$\lambda_1 \neq \lambda_2$, which is a unique signature of parity violation
occurring along with the breakdown of statistical  homogeneity and isotropy.

In light of the signature of stochastic GW background in PTA~\cite{Agazie_2023, EPTA_2023, PPTA_2023}, and proposed
sensitivities of upcoming missions like SKA~\cite{Moore_2014, Janssen_2015} and interferometers such as LISA~\cite{Sathyaprakash_2009, LISA_2017,LISACosmologyWorkingGroup:2023njw}, 
it becomes compelling and feasible to examine the scenario of violation of 
statistical homogeneity, isotropy and parity using the cosmological GW background.
The amplitude and shape of $\ogw$ around its maximum, if compared against a model of 
$\alpha(k)$, can inform us about the level of inhomogeneity of the primordial mean, 
or at least provide an upper bound on it.
The chirality of $\ogw$, if measured in future, shall provide insights into the 
anisotropic structure of primordial mean induced in the tensor perturbations.
The existence of correlation between different modes $\vk_2 \neq \vk_1$ and/or 
correlation between different polarizations of $L$ and $R$ (or $+$ and $\times$), 
shall break degeneracies between this scenario and other possibilities that may 
lead to enhanced tensor power over small scales.
Further, the structure of anisotropies in the $\ogw$ can be readily computed 
in terms of $\beta_{\ell m}$ that captures the anisotropy present in $\alpha(\vk)$
in a fairly generic manner [cf.~Eqs.~\eqref{eq:alpha} and~\eqref{eq:beta}].
Such an exercise, especially for modes with $\ell\neq\ell'$, shall provide 
richer information regarding the precise nature of $\alpha(\vk)$.
We relegate such an analysis to a future work.


\acknowledgements
HVR acknowledges support by the MUR PRIN2022 Project ``BROWSEPOL: Beyond standaRd mOdel With coSmic microwavE background POLarization"-2022EJNZ53 financed by the European Union - Next Generation EU. 
DM thanks Raman Research Institute for support through postdoctoral fellowship.
DM thanks Kinjalk Lochan for helpful discussions and comments.

\appendix
\section{Identifying spectral density in case of inhomogeneities}
\label{app:spectral_density}

We shall outline an alternative method to compute spectral density for a function, in case of statistical inhomogeneity leading to unavoidable spatial
dependence. 
We shall illustrate it with the energy density of the SIGW $\rho_{\rm GW}$.

As discussed in the main text, when the scalar perturbations are inhomogeneous and anisotropic in nature, $\rho_{\rm GW}$ has three components.
Barring prefactors, they are
\begin{eqnarray}
\rho_{\rm GW}(\eta, \vx) &\simeq & f(\eta) + g(\eta,\vx) + h^2(\eta,\vx)\,,
\end{eqnarray}
where the quantities $f,g$ and $h$ are Fourier counterparts of the functions $f(k)$,
$g(\vk_1,\vk_2)$ and $\braket{h_\vk}$ that constitute the two-point correlation
function in Eq.~\eqref{eq:h1h2} as
\begin{align}
  \braket{h^{\lambda_1}_{\bm{k}_1} h^{\lambda_2}_{\bm{k}_2}}
  &= f^{\lambda_1\lambda_2}(k_1)\Dd(\bm{k}_1 + \bm{k}_2) +
  g^{\lambda_1\lambda_2}(\bm{k}_1 , \bm{k_2}) \nn \\
  & + \braket{h^{\lambda_1}_{\bm{k}_1}} \braket{h^{\lambda_2}_{\bm{k}_2}}\,.
\end{align}
We drop the dependence on $\lambda$ for convenience. 
So
\begin{eqnarray}
f(\eta) &=& \iint\f{\d^3 \vk_1\d^3 \vk_2}{(2\pi)^{3}}
f(k_1)\,\delta(\vk_1+\vk_2)\,e^{i\,\vx\cdot(\vk_1+\vk_2)}\,.
\end{eqnarray}
Under suitable change of integration variables, it reduces to
\begin{eqnarray}
f(\eta) &=& \int \f{\d^3 \vk}{(2\pi)^{3}} f(k) 
= \int \d \ln k \f{k^3}{2\pi^{2}} f(k)\,.
\end{eqnarray}
Notice that the quantity $f(\eta)$ is independent of $\vx$ since $f(k)$ is independent
of the wavevector $\hat \vk$ and just a function of $k$. 
Thus the spectral density of $f(\eta)$ can be defined as
\begin{equation}
\f{\d f}{\d \ln k} = \f{k^3}{2\pi^2}f(k)\,.
\label{eq:spec_dens_1}
\end{equation}

Turning to the spectral densities of the other two functions,
\begin{eqnarray}
g(\eta,\vx) &=& \iint\f{\d^3 \vk_1\d^3 \vk_2}{(2\pi)^{3}} 
g(\vk_1,\vk_2)\,e^{i\,\vx\cdot(\vk_1+\vk_2)}\,.
\end{eqnarray}
Again under change of variables, we may rewrite the above as
\begin{eqnarray}
g(\eta,\vx) &=& \int \f{\d^3 \vk}{(2\pi)^3}\,e^{i\vx\cdot\vk}\,\int {\d^3\vq}\,g(\vq,\vk-\vq)\,,\\ 
&=& \int \d\ln k \f{k^3}{(2\pi)^3} \int \d^2\Omega_{\hat \vk} \,e^{i\vx\cdot\vk}
\int {\d^3\vq}\,g(\vq,\vk-\vq)\,. \nn \\
\end{eqnarray}
We can treat the RHS as an inverse Fourier transformation of a function of $\vk$,
that is obtained from integrating over $\vq$. 
In other words, the integral over $\vq$ is the Fourier counterpart of 
$g(\eta,\vx)$.
Thus the spectral density associated with $g(\eta,\vx)$ shall be
\begin{equation}
\f{\d g(\eta, \vx)}{\d \ln k} = \f{k^3}{(2\pi)^3} \int \d^2\Omega_{\hat \vk} \,e^{i\vx\cdot\vk}
\int {\d^3\vq}\,g(\vq,\vk-\vq)\,.
\label{eq:spec_dens_2}
\end{equation}
It depends on $\vx$ because of $g(\vq,\vk-\vq)$ depending on $\hat \vk$.

Similarly, the third term shall be
\begin{eqnarray}
h^2(\eta,\vx) &=& \bigg[ \int\f{\d^3 \vk}{(2\pi)^{3/2}} 
h(\vk)\,e^{i\,\vx\cdot\vk}\bigg]^2\,, \\
&=& \iint \f{\d^3\vk\,\d^3\vq}{(2\pi)^3}h(\vq)h(\vk-\vq)\,e^{i\vx\cdot\vk}\,, \\
&=& \int \f{\d^3 \vk}{(2\pi)^3} \,e^{i\vx\cdot\vk}
\int {\d^3\vq}\,h(\vq)h(\vk-\vq)\,, \\ 
&=& \int \d\ln k \f{k^3}{(2\pi)^3} \int \d^2\Omega_{\hat \vk} \,e^{i\vx\cdot\vk}
\int {\d^3\vq}\,h(\vq)h(\vk-\vq)\,. \nn \\
\end{eqnarray}
This is just an explicit demonstration of product of functions in real space turning out
to be a convolution in Fourier space.
The corresponding spectral density is simply
\begin{equation}
\f{\d h^2(\eta, \vx)}{\d \ln k} = \f{k^3}{(2\pi)^3} \int \d^2\Omega_{\hat \vk} \,e^{i\vx\cdot\vk}
\int {\d^3\vq}\,h(\vq)h(\vk-\vq)\,.
\label{eq:spec_dens_3}
\end{equation}

Adding Eqs.~\eqref{eq:spec_dens_1}~\eqref{eq:spec_dens_2} and~\eqref{eq:spec_dens_3},
we get the spectral density associated with $\rho_{\rm GW}(\eta, \vx)$ to be
\begin{eqnarray}
\f{\d\rho_{\rm GW}(\eta,\vx)}{\d\ln k} &=& \f{k^3}{2\pi^2}
\bigg[ f(k) + \f{1}{4\pi} \int \d^2\Omega_{\hat \vk} \,e^{i\vx\cdot\vk} \nn \\
& & \times \int {\d^3\vq}\,[ g(\vq,\vk-\vq) + h(\vq)h(\vk-\vq)] \bigg]\,. \nn \\
\end{eqnarray}
This method may be less easier for computational purpose than the method discussed 
in the main text [cf.~Eqs.~\eqref{eq:phk-k} and~\eqref{eq:phk1k2}].
But the advantage of this way of expressing spectral density is that, we do not
rely on the smallness parameter $\Delta$.
Further, we can readily extract the components of monopole and the higher-order 
multipoles of the anisotropic structure of $\rho_{\rm GW}$ by expanding 
$e^{i\,\vx\cdot\vk}$ present in the angular integral, in terms of spherical 
harmonics.

%


\end{document}